# Transition Temperature and Upper Critical Field in SmFeAsO$_{1-x}$F$_x$ Synthesized at Low Heating Temperatures

S. J. Singh, J. Shimoyama, A. Yamamoto, H. Ogino and K. Kishio

*Abstract*–: **Low-temperature synthesis is a promising and potentially effective method for improving superconducting properties. We report on the fabrication of polycrystalline samples of SmFeAsO$_{1-x}$F$_x$ with nominal $x$ content varying in a wide range of $x$ = 0–0.35 synthesized at 900 °C. This synthesis temperature is around 300 °C lower than the conventional synthesis temperature. The variation in the lattice parameters and transition temperature ($T_c$) of various F-doped samples indicates that reduction of the unit cell volume ($V$) seems to be the main reason for the rise of $T_c$ up to 57.8 K. Magnetoresistance measurements showed that the upper critical field slope (d$H_{c2}$/dT) increased with increasing F concentration up to $x$ = 0.2, where it reached a maximum value of -8 T/K corresponding to a coherence length ($\xi_{GL}$) of 10 Å. At still higher F doping levels, d$H_{c2}$/dT and the low field $J_c$ decreased; above 0.5 T, however, $J_c$ had almost the same value. Compared with previous reports, the present synthesis route with low synthesis temperatures and commonly available FeF$_2$ as the source of F is more effective at introducing F into the SmFeAsO system and thereby resulting in improved superconducting properties for the system. In addition, this new sample preparation method also reduces unnecessary problems such as the evaporation of F and reaction between the crucible and superconductor during the solid-state reaction.**

## I. INTRODUCTION

THE RECENT discovery of a new class of high-temperature superconductors REFeAsO (1111-type; RE = rare earth) [1] with a high critical temperature ($T_c$) and high upper critical field ($H_{c2}$) has attracted interest worldwide [2]. Among the pnictides known to date, SmFeAsO$_{1-x}$F$_x$ (Sm 1111) superconductors are considered to be very promising owing to their high $T_c$ of up to 55 K under optimal doping ($x$ = 0.2) [3]–[4]. The parent compound SmFeAsO is a normal semimetal, and superconductivity is induced in the compound by either replacing O$^{-2}$ ions with F$^-$ ions or forming oxygen vacancies [2], [5]. The properties of this new class of superconductors can be well explained by characterization of high-purity (pure phase) samples of the compound. Achieving high-phase purity polycrystalline oxypnictides (1111) is a major challenge for iron-based superconductors. Since their discovery, several studies have focused, in particular, on the synthesis conditions of F doping in SmFeAsO$_{1-x}$F$_x$. These studies showed that a high pressure (~6 GPa) [3] or high temperature (conventionally ~1200 °C) [6]–[7] is suitable for introducing F into SmFeAsO. However, the vast majority of F-doping experiments to date have been performed at ambient pressure [8]–[13] with the samples sealed in a quartz tube. In such cases, the necessary high sample preparation temperature of about 1200 °C is very close to the temperature limit of the quartz tube employed in the solid-state method at ambient pressure. Therefore, it is very likely that the quartz tube cracks during the preservation or cooling process and thus leading to failure of the synthesis experiments. This is an incontrovertible problem with the solid-state preparation process.

Another problem encountered during synthesis is the differences between the real and nominal compositions of the F-doped system. Since F is a light element, the high synthesis temperature for oxypnictides can induce a decrease in the real F content and reduce the superconducting properties. Therefore, to overcome this problem, a new low-temperature synthesis route is urgently needed. To date, only a few reports [8]–[10] on the bulk fabrication of SmFeAsO$_{1-x}$F$_x$ at 850–900 °C are available; a recent report [8] used a two-step process, which resulted in an onset $T_c$ of 55 K for $x$ = 0.4. In this study, a series of polycrystalline SmFeAsO$_{1-x}$F$_x$ samples ($x$ = 0–0.35) were prepared at 900 °C, which is far below the conventional synthesis temperature. This further proves that a single-step bulk fabrication process should be possible at similar or even lower temperatures. We investigated our samples by measuring the X-ray powder diffraction (XRD), resistivity, and magnetization, the results clearly revealed the superconducting properties of the obtained compounds. The primary aim of this work was to develop an efficient F-doping process with low synthesis temperatures. Our results demonstrated improved superconductivity of Sm 1111 with a maximum onset $T_c$ of 57.8 K and the highest upper critical field slop of -8 T/K. The present synthetic method reduces the formation of impurity phases, especially those of SmOF, and seems to be better suited for F-doping than previously reported methods [8]–[13].

Manuscript received 08 October 2012. SJS is supported by the Japan Society for the Promotion of Science (JSPS), Government of Japan. This work was partially supported by the Japan Science and Technology Agency (JST), PRESTO, SICORP, and TRIP.

The authors are at the Department of Applied Chemistry, The University of Tokyo, 7-3-1 Hongo, Bunkyo-ku, Tokyo 113-8656 Japan (Corresponding author phone: +81-3-5841-7766; fax: +81-3-5689-0574; e-mail: shivjees@gmail.com ).

## II. EXPERIMENT

Polycrystalline samples with nominal compositions of SmFeAsO$_{1-x}$F$_x$ ($x = 0$–0.35) were synthesized by a single-step solid-state reaction method using high-purity (>99.99%) SmAs, Fe$_2$O$_3$, FeF$_2$, and Fe as starting materials. The raw materials were mixed according to the stoichiometric ratios, pressed into pellets, and then sealed in evacuated quartz ampoules. Ta foils were used to hold the pellets and avoid contact with the quartz. Optimally doped samples ($x = 0.2$) were synthesized in a very wide temperature range 800–1100 °C in steps of 50 °C (not shown here) in order to find the best synthesis temperature (900 °C). Finally, the quartz ampoules, with all samples of SmFeAsO$_{1-x}$F$_x$ inside, were heated to 900 °C for 45 h followed by cooling to room temperature at a rate of 30 °C/h. All chemical manipulations were performed in an argon-filled glove box. Several samples belonging to different batches were prepared by this procedure, and they showed high reproducibility with regard to the superconducting properties. The resulting samples were characterized by powder XRD using Cu-$K_\alpha$ radiation. The lattice parameters were obtained by a least squares fit of the observed $d$ values.

The temperature dependence of the resistivity in magnetic fields of up to 9 T was measured by a standard four-probe method with an ac current frequency of 15 Hz using a commercial measurement system (Quantum Design PPMS Model 6000). Magnetic measurements were performed with a SQUID magnetometer (Quantum Design: MPMS-XL5s). The measurements were carried out in a magnetic field upon heating after zero-field cooling (ZFC) and, subsequently, field cooling (FC).

## III. RESULTS AND DISCUSSION

Fig. 1 shows the powder XRD patterns of representative samples of SmFeAsO$_{1-x}$F$_x$ ($x = 0$, 0.1, 0.2, 0.25, 0.3, and 0.35). The majority of the observed reflections could be satisfactorily indexed on the basis of the tetragonal ZrCuSiAs-type structure [14]. Minor amounts of SmAs, SmOF, FeAs, and Sm$_2$O$_3$ were observed as impurity phases for samples with $x > 0.2$.

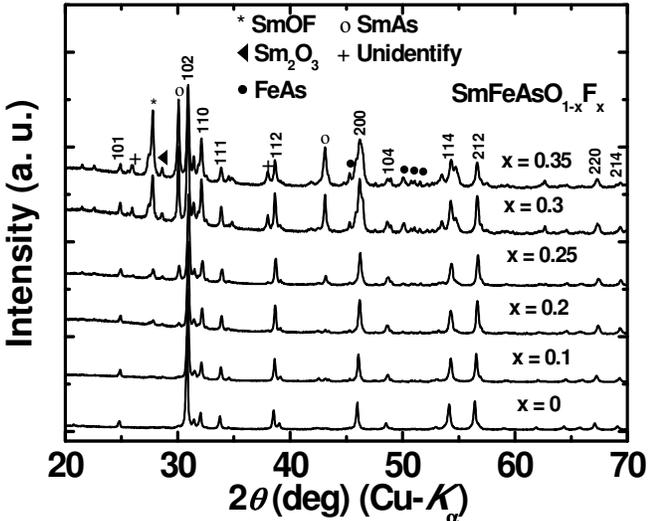

Fig. 1. Powder X-ray diffraction patterns of bulk SmFeAsO$_{1-x}$F$_x$ with various nominal F -content ($x$) heated at 900 °C for 45 h.

The zero-field resistivity as a function of the temperature for the nominal compositions of SmFeAsO$_{1-x}$F$_x$ ($x = 0$–0.3) is shown in Fig. 2. Pure undoped SmFeAsO showed a broad anomaly in the resistivity at a temperature of around 150 K owing to the collective effect of a spin density wave (SDW) instability and structural phase transition from tetragonal to orthorhombic symmetry [14]-[15]. Doping of F suppressed the SDW transition, and the resistivity decreased monotonously with decreasing temperature. The onset of the superconducting transition occurred at 44.5 and 57.1 K for $x = 0.1$ and 0.2, respectively. The decrease in the resistivity suggests that the charge carrier density increased. Upon further increase in the F content, $T_c$ reached its highest onset value of 57.8 K, which is the same value as reported for thin-film Sm 1111 [16]. The samples with $x = 0.3$ and 0.35 showed approximately constant transition temperatures of 57.7 K. The criteria used to determine $T_c$ are shown in the inset of Fig. 2; they are similar to the methods in earlier reports [17]. The residual resistivity ratios ($RRR = \rho_{300K}/\rho_{58K}$) of 5.1, 8.0, and 7.8 for the samples with $x = 0.15$, 0.2, and 0.25, respectively, suggests good metallic normal-state connectivity.

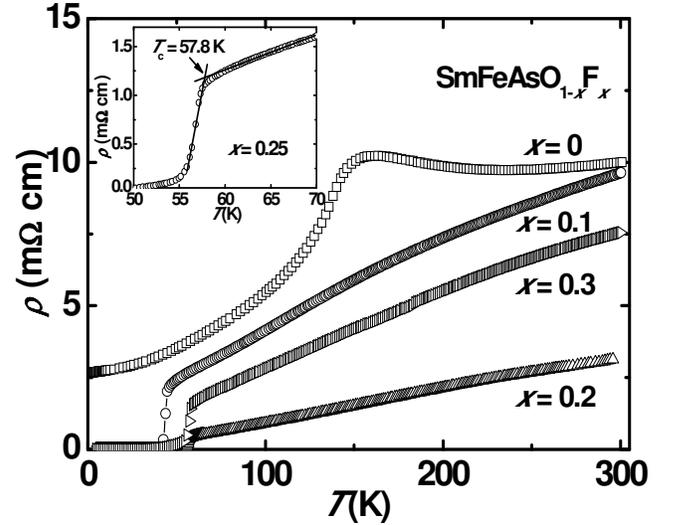

Fig. 2. Plot of the resistivity ($\rho$) versus temperature for bulk SmFeAsO$_{1-x}$F$_x$. The inset shows the criterion to determine $T_c$.

Fig. 3 shows the variations in the lattice parameter $a$, unit cell volume $V$, and transition temperature $T_c$ with a F content of $x$. The refined lattice parameters of the doped samples were smaller than those of the parent compound ($a = 3.939(1)$ Å and $c = 8.506(2)$ Å) and thus indicate successful doping of F at O sites. The present data were also compared with the results of Liu $et\ al.$ [15] and Chen $et\ al.$ [11], for which samples of SmFeAsO$_{1-x}$F$_x$ were synthesized at conventional synthesis temperatures (~1200 °C). Both lattice parameters $a$ and $c$ decreased upon an increase in F content of up to $x = 0.2$ owing to the smaller ionic size of F$^-$ compared to O$^{2-}$. Further increases in the nominal F content ($x > 0.2$) caused the lattice parameter $a$ to shrink continuously (Fig. 3), whereas $c$ seemed to increase slightly; however, the cell volume $V$ remained almost constant, indicating the solubility limit of F ions at the O sites. This means that a solid solution of F in SmFeAsO$_{1-x}$F$_x$ is valid in the limited range of $0 \leq x \leq 0.2$. For $x > 0.2$, excess F mainly formed a secondary phase of SmOF as confirmed by XRD (Fig. 1). Because of the presence of such impurity phases, the resistivity value also started to increase for higher levels of F-doping ($x > 0.2$). The upper panel of Fig. 3 shows the dependence of the onset $T_c$ upon the F content ($x$). $T_c$ reached a maximum value of 57.8 K for $x = 0.25$ but became constant at higher doping levels. Interestingly, the onset $T_c$ of 57.8 K of the present study was 2 or 3 K higher than the values reported earlier [8]–[13], [15] for bulk SmFeAsO$_{1-x}$F$_x$. The increase in $T_c$ can be attributed to the shrinkage of the crystal lattice, which caused stronger chemical pressure on the FeAs plane [18] and was induced by an increase in the actual F-doping level under present synthesis.

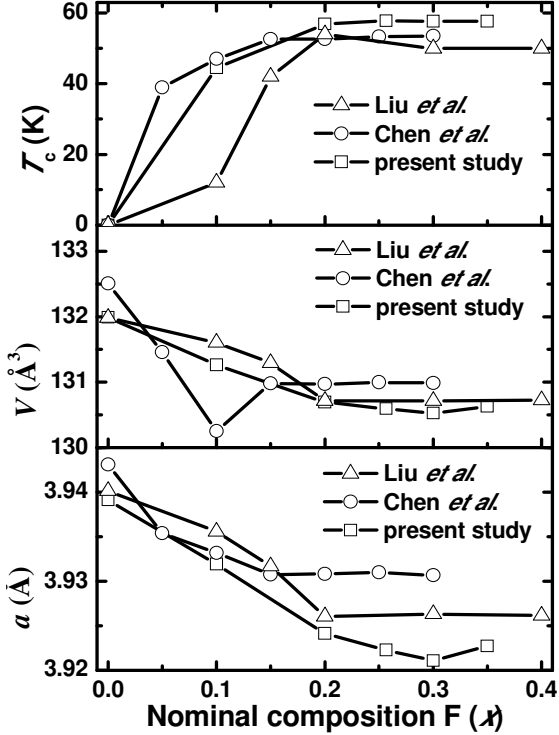

Fig. 3. Variations in the lattice constant $a$, cell volume $V$, and transition temperature $T_c$ with nominal F content compared with the results of Liu *et al.* [15] and Chen *et al.* [11], for which the samples were synthesized at a high temperature of ~1200 °C.

The lattice parameter $a$ and cell volume were always smaller than those reported by Liu *et al.* and Chen *et al.* for the same nominal compositions. Thus, the present low-temperature synthesis using FeF$_2$ seems to be an effective method for introducing F into bulk SmFeAsO as it resulted in the highest $T_c$ reported so far.

In the previous report on low-temperature synthesis of F-doped SmFeAsO [8], a two-step process was applied at a temperature of 850 °C. A maximum $T_c$ of 55 K was observed for $x = 0.4$ with a large amount of the SmOF impurity phase, which clearly indicates that part of F did not dissolve in the lattice. In another report [9], samples of SmFeAsO$_{0.8}$F$_{0.2}$ also had large impurity phases; therefore, no systematic variation in the lattice parameters could be observed. Compared with these results, the present samples had smaller lattice constants and a higher $T_c$ even at low F concentrations. The fraction of impurity phases is much smaller. These results confirmed the relatively low volatilization of F and an efficient F-doping effect of the present synthesis process.

The temperature dependence of ZFC and FC magnetization is shown in Fig. 4. The value $|4\pi M/H|$ was larger than unity because of the demagnetization effect of the bulk sample. The diamagnetic transition of the sample with $x = 0.1$ was sharp and well-defined, indicating a homogeneous F concentration and relatively strong grain coupling. Upon further doping

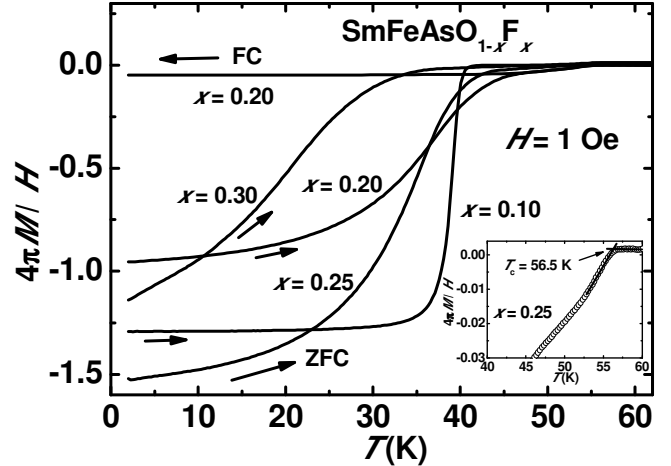

Fig. 4. Temperature dependence of ZFC and FC magnetization in various SmFeAsO$_{1-x}$F$_x$ bulk samples. The inset shows ZFC close to $T_c$.

($x > 0.10$), the ZFC curves clearly show two distinct steps: the step near $T_c$ is due to intragrain shielding, while that at the lower temperature reflects intergrain shielding. This double-step is a known feature of the granular magnetic response and is similar to the results of previous reports [19], [20]. Inhomogeneous superconductors are also known to show such features, as discussed in [21]. Hence, the double step feature in the observed extended diamagnetic transition can be considered to be a combined effect of the presence of secondary phases and the electromagnetic granularity of the sample. The difference in $T_c$ (onset) between resistivity and magnetization measurements was observed to be ~1 K in these superconducting materials.

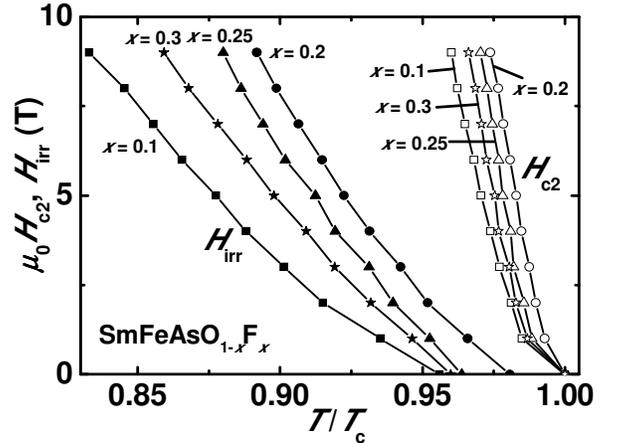

Fig. 5. Temperature dependence of the upper critical field ($H_{c2}$) and irreversibility field ($H_{irr}$).

To obtain information about the upper critical field ($H_{c2}$) and flux pinning properties, the temperature dependence of the resistivity was measured under various magnetic fields of up to 9 T. As the field was increased, $T_c$ (onset) slightly decreased, while the zero resistance temperature largely decreased owing to a significant broadening of the resistivity transition. The upper critical field $H_{c2}$ and irreversibility field $H_{irr}$, as determined by the 90% and 10% values of the normal state resistivity ($\rho_n$), are summarized in Fig. 5. The slopes of the $\mu_0 dH_{c2}/dT$ is estimated between 1 T and 9 T fields from $H$- $T$ diagram and the values are -6.8, -8.0, -7.4, and -6.6 T/K for samples with $x$ = 0.1, 0.2, 0.25, and 0.3, respectively, which agrees with the magnetoresistivity data of the same class of compounds [22]-[24] even though much lower values were reported from calorimetric measurements on single crystals of Nd 1111 superconductors [25]. Such high $\mu_0 dH_{c2}/dT$ values have already been reported [20], [22], [26] for this system processed around the conventional synthesis temperature, but we were able to achieve the same/very close values at 900 °C with improved $T_c$. Furthermore, under the simplified assumption of single-band superconductivity and using the Werthamer–Helfand–Hohenberg (WHH) formula $H_{c2}(0) = -0.693T_c \, (dH_{c2}/dT)$ [27] which is based on the orbital-paramagnetic effect, it is possible to estimate $\mu_0 H_{c2}(0) = 210$, 315, 298, and 259 T for $x$ = 0.1, 0.2, 0.25, and 0.3, respectively, and thus exceeding the BCS paramagnetic limit $H_p = 1.84T_c$. The Ginzburg–Landau coherence length at zero temperature $\xi_{GL} = (\Phi_0/4\pi H_{c2})^{1/2}$, where $\Phi_0$ is the flux quantum equal to $2.07 \times 10^{-7}$ G cm$^2$, was 12.5, 10.2, 10.5, and 11.3 Å for $x$ = 0.1, 0.2, 0.25, and 0.3, respectively. $H_{c2}$ of oxypnictides can be described well by a theory for the multiband $s^\pm$ pairing in the clean limit [28], which suggests the Fulde–Ferrel–Larkin–Ovchinnikov (FFLO) state. The actual $H_{c2}$ of real materials is generally influenced by the interplay of orbital and spin-paramagnetic pairs breaking in different bands. The relative effect of both the orbital and spin (Pauli) paramagnetic effects are explained in terms of the Maki parameter [28], [29] $\alpha = 1.44 H_{c2}^{orbital}(0)/H_p(0)$. The calculated $H_{c2}$ (by WHH) for the present samples were much higher than the BCS paramagnetic limit. In this situation the Maki parameter ($\alpha$) will be greater than 1 which suggests that Sm 1111 might be affected by strong Pauli limiting of $H_{c2}$, which is likely to form the FFLO ground state. In multiband superconductors the orbital pair breaking and FFLO instability can be tuned by doping [28], and the effect of the doping in tuning of $H_{c2}(T)$ is due to the change in the Fermi velocity that in clean limit determines $\alpha$ [28], [30]. Hence, spin paramagnetic pair-breaking might be a dominant mechanism in REFeAsO. The calculated $H_{c2}(0)$ value from high-field measurements of up to 85 T [26], [30]–[31] has been reported to be inconsistent with the predicted result of WHH theory without the spin paramagnetic effect.

The hysteresis loop at 5 K for $x$ = 0.2, 0.25 and 0.3 is presented in the inset of Fig. 6. The samples were of rectangular shape, and the magnetic moment ($m$) was normalized by the weight of the specimen. A very small paramagnetic background was observed in the samples with $x \leq 0.2$, whereas a relatively large paramagnetic background appeared in samples with larger $x$. The background seemed to rise with an increase in impurity phases. The width of the hysteresis loop $\Delta m(H)$ is small, which is a general feature of oxypnictides and could be due to either poor intergrain connectivity or weak intragranular pinning [32]–[33]. We calculated the critical current density $J_c$ for our samples on the basis of $J_c = 20\Delta m/Va(1-a/3b)$, where $a$ and $b$ are the lengths of the shorter and the longer edge, respectively, $V$ is the volume of the sample, and $\Delta m$ is the hysteresis loop width. The magnetic field dependence of the critical current density ($J_c$) at 5 K of all samples is compared in Fig. 6. Note that, as the F content increased, the $J_c$ value was enhanced to reach a maximum value for $x$ = 0.2 at the low field region (H $\leq$ 0.5 T). However, at the high field region, the calculated $J_c$ value for x $\geq$ 0.2 did not vary much and revealed a weak dependence on external fields compared to $x < 0.2$. For $x = 0.2$, the estimated $J_c$ was ~$1.5 \times 10^3$ A/cm$^2$ at $T = 5$ K and $H = 0.2$ T.

The relatively smaller amount of impurity phases, larger decrease in the lattice parameters, and higher $T_c$ and $H_{c2}$ values suggest that the present synthesizing process is more efficient at introducing F into the SmFeAsO compounds than conventional methods [8]–[13], [15]. The present process uses FeF$_2$ (melting point = 970 °C) in place of the commonly used SmF$_3$ (melting point = 1306 °C) as the starting material. The lower heating temperature improves the activity of the fluoride and makes reactions between fluoride and other raw materials much easier. Accordingly, reactions between F and the quartz tube are clearly diminished. The lower synthesis temperature also suppresses the formation of the SmOF impurity phase. This relatively low-temperature method could become a promising technique for the preparation of high-quality F-doped REFeAsO polycrystalline samples.

## IV. CONCLUSION

The effect of F doping on the superconducting properties of SmFeAsO$_{1-x}$F$_x$ ($x = 0$–0.35) using a low-temperature synthesis was systematically studied. The results showed that good quality samples can be prepared at relatively low temperatures of 900 °C using FeF$_2$ as the source of F. A study of the lattice constants, $T_c$, and resistivity ($\rho$) revealed that $T_c$ increased with reduced unit cell volume ($V$), while the resistivity reached a minimum value with optimal F doping. The present synthesis route is more effective than conventional methods at introducing F in SmFeAsO, reducing the impurity phases, raising $T_c$ up to 57.8 K, and increasing the $H_{c2}$ slope to -8 T/K. This synthesis route is also good for a sample preparation employing sealed quartz tubes that can be maintained at 900 °C for a long time. We believe this method will enable further exploration of Sm-based oxypnictide materials to achieve additional improvements in the superconducting properties and the development of other applications, especially with respect to superconducting wires.

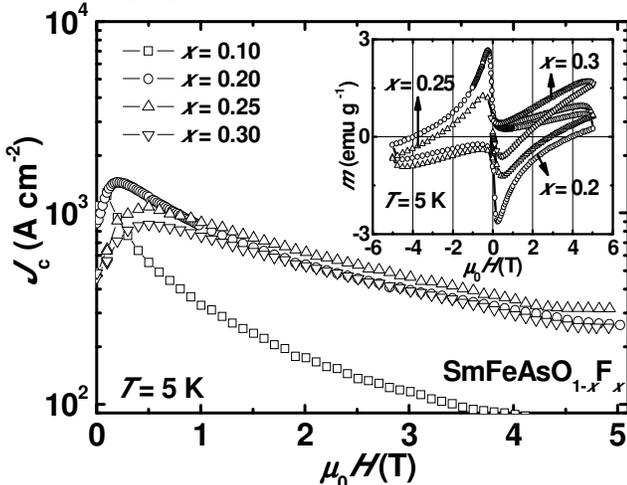

Fig. 6. Magnetic field dependence of the critical current density ($J_c$) for bulk SmFeAsO$_{1-x}$F$_x$. The inset of the figure shows the magnetization ($m$) hysteresis loop at 5 K for $x = 0.2$–0.3.